\documentclass[a4paper]{article}

\usepackage{INTERSPEECH2022}
\usepackage{multirow}
\usepackage{amsmath}
\usepackage{graphicx}

\title{A Graph Isomorphism Network with Weighted Multiple Aggregators for Speech Emotion Recognition}
\name{Ying Hu{$^1$ $^2$}, Yuwu Tang{$^1$ $^2$}, Hao Huang{$^1$ $^2$}, and Liang He{$^2$ $^3$}}
\address{
  $^1$Key Laboratory of signal detection and processing in Xinjiang, China\\
  $^2$School of Information Science and Engineering, Xinjiang University, Urumqi, China\\
$^3$Department of Electronic Engineering, Tsinghua University, China}
\email{huying@xju.edu.cn, yuwu\underline{~}tang@stu.xju.edu.cn}

\begin{document}

\maketitle
\begin{abstract}
Speech emotion recognition (SER) is an essential part of human-computer interaction. In this paper, we propose an SER network based on a Graph Isomorphism Network with Weighted Multiple Aggregators (WMA-GIN), which can effectively handle the problem of information confusion when neighbour nodes' features are aggregated together in GIN structure. Moreover, a Full-Adjacent (FA) layer is adopted for alleviating the \(over$-$squashing\) problem, which is existed in all Graph Neural Network (GNN) structures, including GIN. Furthermore, a multi-phase attention mechanism and multi-loss training strategy are employed to avoid missing the useful emotional information in the stacked WMA-GIN layers. We evaluated the performance of our proposed WMA-GIN on the popular IEMOCAP dataset. The experimental results show that WMA-GIN outperforms other GNN-based methods and is comparable to some advanced non-graph-based methods by achieving 72.48\% of weighted accuracy (WA) and 67.72\% of unweighted accuracy (UA).
\end{abstract}


\begin{figure*}[htbp]
	\centering
	\includegraphics[width=16cm]{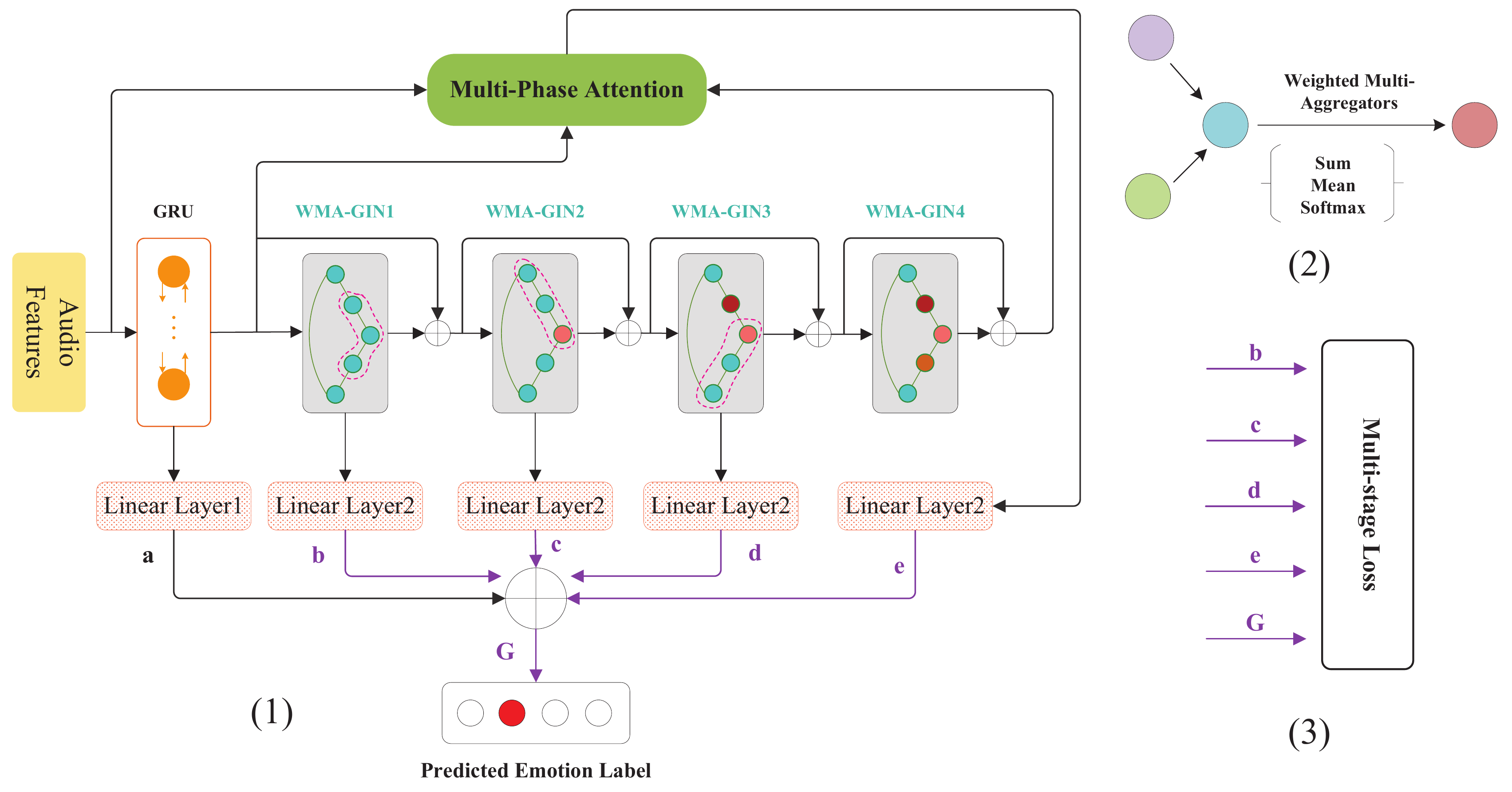}
	\caption{Illustration of our proposed SER method. (1)The overview of the WMA-GIN SER network. (2) Weighted Multiple Aggregators. (3) Multi-stage loss training strategy.}
	\label{proposed network}
\end{figure*}
\section{Introduction}

Speech Emotion Recognition (SER) task is a well-studied field in the domain of affective computing and has been essential for computers understanding human's mood state \cite{anagnostopoulos2015features, AKCAY202056}. The SER has attracted many researchers' attention and been applied to plenty of fields such as automated customer service systems, smart voice assistants, human psychotherapy, and so on \cite{poria2017review, yeh2019interaction}. However, SER is an extraordinarily complicated task even for human beings because speech emotion is elusive.

Graph Neural Network (GNN) has been an active research field for the last ten years and made significant advancements in graph representation learning \cite{scarselli2008graph, bronstein2017geometric, battaglia2018relational}. GNNs broadly follow a recursive neighborhood aggregation (or message passing) scheme, where each node aggregates the feature vectors of its neighbors to compute its new feature vector \cite{xu2018representation, gilmer2017neural}. In recent years, some methods based on GNN have been applied in audio processing fields, including few-shot audio classification \cite{zhang2019few}, anti-spoofing \cite{tak2021graph} and so on. Su et al. \cite{su2020improving} proposed a framework in imposing a graph attention mechanism on a gated recurrent unit network (GA-GRU) to improve utterance-based SER. Shirian et al. \cite{shirian2021compact} constructed a compact Graph Convolution Network architecture which for the first time only utilized GNN for SER task. Liu et al. \cite{liu2021graph} proposed a novel LSTM-GIN model, which applies Graph Isomorphism Network (GIN) \cite{xu2018powerful} on LSTM outputs for global emotion modeling in the non-Euclidean space. These SER methods based on GNN all model the time sequence features of speech signals as graphs and treat the SER as a graph classification task. 

There are many GNN variants with different neighborhood aggregation, such as GCN \cite{kipf2017semi}, GAT \cite{velivckovic2018graph}, PATCHY-Diff \cite{ying2018hierarchical}. In this paper, we choose GIN as the basic GNN structure because it possesses the discriminative power over other GNNs \cite{xu2018powerful}. Nevertheless, in the abovementioned papers, GNNs all adopt a single aggregator in a GNN layer, which does not extract enough information from the nodes’ neighbourhoods and thus limit their expressive power and learning abilities. Corso et al. \cite{corso2020principal} have mathematically proved the necessity of adopting multiple aggregators. We adopt weighted multiple aggregations in the GIN structure and choose three kinds of aggregators: sum, mean and softmax aggregations. We refer to our proposed GNN-based SER structure as Weighted Multiple Aggregators GIN (WMA-GIN). 

Many research fields utilizing GNN, including SER, require the interaction between nodes that are not directly connected. They achieve this by stacking multiple GNN layers for long-range sequence problems. However, as the number of layers increases, the number of nodes in each node’s receptive field grows exponentially. This phenomenon is named as \(over$-$squashing\) by U. Alon and E. Yahav \cite{alon2020bottleneck}. For alleviating this problem in our network, we transform the neighbourhood relationship to the Full-Adjacent (FA) at one layer of the WMA-GIN. Furthermore, we have made the ablation experiments to demonstrate which layer of the WMA-GIN should be introduced into the FA for the SER task. 

A Bi-GRU layer extracts global context sequence information before the stacked WMA-GIN layers. Moreover, multi-phase attention (MPA) is adopted to extract emotional information in underutilization. And a multi-stage loss training strategy is employed the handle the output information from different stages in our network. The experimental results show the effectiveness of our proposed method.

The contributions of this paper are summarized as follows:

(1) We propose a Weighted Multiple Aggregators GIN (WMA-GIN) for the SER task, and the experimental results show its effectiveness for obtaining more accurate information from neighbour nodes.

(2) We introduce a Full-Adjacent (FA) layer into the WMA-GIN for relieving the \(over$-$squashing\) problem. And we have explored which layer of WMA-GIN is more suitable to adopt the FA for the SER task.

(3) Our proposed WMA-GIN achieves 72.48\% of WA and 67.72\% of UA on the IEMOCAP dataset, outperforming other state-of-the-art GNN-based SER methods.

\section{Proposed Method}

In this section, we will introduce our proposed WMA-GIN network architecture, the graph construction from the audio utterances, GIN, and weighted multiple aggregators. At last, we discuss the Fully-Adjacent layer and multi-stage loss training strategy.  

\subsection{WMA-GIN Architecture}

The architecture of WMA-GIN is illustrated in Fig.~\ref{proposed network}. A Bi-GRU layer is used for extracting the global context information of the features among long-range time series, and the hidden dimension of each direction is 128. Then, the output \(\mathbf{g}\) of Bi-GRU is fed into stacked WMA-GIN layers for extracting the features of higher resolution and learning the information with richer emotional characteristics. The number of WMA-GIN layers is set to 4. There are residual connections between each two adjacent WMA-GIN layers. Moreover, multi-phase attention (MPA) is employed for extracting emotional features that might be omitted from different phases of the network. Similar to self-attention \cite{vaswani2017}, each of three linear projections is performed on the input features, the output of Bi-GRU, and output of the last WMA-GIN layer, respectively, such that they are transformed as \(query\), \(key\) and \(value\). The output dimension of the linear projection is equal to the hidden size of the WMA-GIN layer.

Finally, the outputs of the Bi-GRU layer, WMA-GIN1, WMA-GIN2, WMA-GIN3, and MPA are fed into the corresponding linear layer to produce the predicted results \(a\), \(b\), \(c\) \(d\), \(e\), respectively. As shown in Figure~\ref{proposed network}, all outputs of predicted results are summed up as \(G\). Then, the predicted emotion label is obtained via a softmax layer. 
\\

\noindent \textbf{Graph Construction}

Following the prior works \cite{shirian2021compact,liu2021graph}, we adopt frame-to-node transformation for our graph construction. The node feature \(\mathbf{s}_{i} \in \mathbb{R}^{h}\) is obtained from the corresponding audio frame with the feature dimension of \(h\). A graph \(\mathcal{G}=(\mathcal{V}, \mathcal{E})\), where \(\mathcal{V} \in\left\{v_{i}\right\}_{i=1}^{n}\) is the set of \(n\) nodes, and \(\mathcal{E}\) is the set of all edges of each nodes. So the feature matrix of each graph is denoted as \(\mathbf{S} \in \mathbb{R}^{n \times h}\). The adjacency matrix of \(\mathcal{G}\) is denoted by \(\mathbf{A} \in \mathbb{R}^{n \times n}\) where an element \((\mathbf{A})_{i j}\) denotes the edge weight connecting node \(v_{i}\) and \(v_{j}\). Here, each node has two neighboring nodes corresponding to the previous and following one frame. The relationship of node neighborhood is defined as a cycle construction where the node of the first time frame is connected with the last time frame.\\

\noindent \textbf{Graph Isomorphism Network (GIN)}

The basic GIN is a GNN architecture proposed by \cite{xu2018powerful} that is provably the most expressive among the class of GNNs and is as powerful as the Weisfeiler-Lehman graph isomorphism test. The GIN has achieved excellent performance for graph classification and node classification tasks. The GIN also follows a neighborhood aggregation scheme, where the representation vector of a node is computed by recursively aggregating and transforming representation vectors of its neighboring nodes \cite{xu2018powerful}. The original GIN adopts a sum aggregation strategy for aggregating information representation from the neighboring nodes. The calculation of sum aggregation is as follow:
\begin{equation}
	sum _{i}\left(X^{l}\right)=\sum_{j \in N(i)} X_{j}^{l}				
\end{equation}
where \(X^{l}\) denotes the nodes’ features at layer \(l\), and \(N(i)\) the neighbourhood of node \(i\).

\subsection{Weighted Multiple Aggregators (WMA)}

While a single aggregator is not enough for differentiating between neighbourhood messages. For example, a node of graph \(a\) receives two messages \($\{$0.2, 0.2$\}$\) from two neighbour nodes. And the received messages of another node of graph \(b\) are \($\{$0, 0.4$\}$\). In this case, the sum aggregator cannot distinguish between graph \(a\) and graph \(b\). For alleviating the abovementioned information confusion, we introduce two other aggregators: mean aggregator and softmax aggregator. Their calculations are as follows:
\begin{equation}
	mean_{i}\left(X^{l}\right)=\frac{1}{d_{i}} \sum_{j \in N(i)} X_{j}^{l}				
\end{equation}
\begin{equation}
	softmax_{i}\left(X^{l}\right)=\sum_{j \in N(i)} \frac{X_{j}^{l} \exp \left(X_{j}^{l}\right)}{\sum_{k \in N(i)} \exp \left(X_{k}^{l}\right)}				
\end{equation}
where \(d_{i}=|N(i)|\) which denotes the number of neighbour nodes. The messages from different nodes can be assigned different weights through softmax aggregation. During the model training, the information will constantly flow to the part with salient emotional features. As the graph structure in our proposed SER method is relatively simple where each node gets two neighbour nodes, we do not choose to scale the results of aggregations as \cite{corso2020principal} does. Finally, the node representation in WMA-GIN updates as:
\begin{equation}
	\begin{aligned}
	&X_{i}^{(l)}=\operatorname{MLP}^{(l)}( ( 1+\epsilon^{(l)} ) \cdot X_{i}^{(l-1)}\\
	&+\alpha \sum_{j \in \mathcal{N}(i)} X_{j}^{(l-1)}+\beta \frac{1}{d_{i}} \sum_{j \in N(i)} X_{j}^{(l-1)}\\&+\gamma \sum_{j \in N(i)} \frac{X_{j}^{(l-1)} \exp X_{j}^{(l-1)}}{\sum_{k \in N(i)} \exp X_{k}^{(l-1)}})
	\end{aligned}				
\end{equation}
where MLP denotes Multilayer Perceptron that is a linear layer, \(X_{i}^{(l)}\) denotes feature representation of node \(i\) in the \(l-th\) hidden layer. And \(X_{i}^{(0)}\) is initialized as \(g_{i}\) which denotes the output of Bi-GRU layer. \(\epsilon\) is a learnable parameter. \(\alpha\), \(\beta\), \(\gamma\) are three hyperparameters of weights for three aggregators.  

\subsection{Fully-Adjacent layer}

To ease the \(over$-$squashing\) problem in all GNNs, we modified the second WMA-GIN layer to be a Fully-Adjacent (FA) layer. The FA layer is a WMA-GIN layer in which every pair of nodes is connected by an edge. This does not change the type of WMA-GIN layer nor add trainable parameters but only changes the adjacency relationship of the nodes in a single layer. After adding the FA layer, only the second WMA-GIN layer allows the topology-aware node-representations to interact directly and consider nodes far beyond their original neighbors \cite{alon2020bottleneck}.

\subsection{Multi-stage Loss}

A multi-stage loss training strategy is adopted in our proposed SER method. The outputs of four WMA-GIN layers and the final output are defined as five different stages in the SER network. We denote them as \(b\), \(c\), \(d\), \(e\), \(G\) in Fig.~\ref{proposed network}. Generally, the stage deeper, the obtained emotional information more salient. Thus the output \(G\) is assigned with the larger weights. Cross-entropy (CE) loss is used as the loss function. Note that the multi-stage loss training strategy is only used during the training phase. The calculation of the final loss function is as follows:
\begin{equation}
Loss=\frac{\sum_{i=1}^{I} i \cdot L_{CE}}{\sum_{i=1}^{I} i}
\end{equation}

where \(i/\sum_{i=1}^{I} i\) is the weight of loss at the \(i-th\) stage and \(I\) is the number of total stages.

\section{Experiments}

\subsection{Dataset}
We performed the experiments on the Interactive Emotional Dyadic Motion Capture (IEMOCAP) dataset \cite{busso2008}, which is a popular benchmark dataset for emotion recognition. The corpus contains approximately 12 hours of data over five dyadic sessions with ten subjects. Each interaction conversation is around 5 minutes long and segmented into multiple sentences. The single categorical emotional label was assigned to every utterance, with over two of three annotators agreeing on the emotional labels. We performed the experiments on the task of four-class emotion classification for a fair comparison with other methods. The samples contain 5531 utterances (1,103 angry, 1636 happy merged with excitement, 1,084 sad, 1,708 neutral).

\subsection{Node features}

Follow two prior work \cite{liu2021graph, su2020improving}, we extract 78-dimensional frame-level Low Level Descriptors (LLDs) from \cite{schuller2010interspeech} using the openSMILE toolkit \cite{10.1145/1873951.1874246}. We also make the experiments using 128-dimensional log-Mel spectrograms as \cite{zhong2020lightweight} does. For each sample, we use a sliding window of length 25ms (with a stride length of 10ms) to extract the LLDs locally. We set each graph length to 120 as \cite{shirian2021compact} and \cite{liu2021graph} do, that means each graph contains 120 nodes. The graph label is the same as its original utterance. An utterance may be cut as several graphs because the lengths of some utterances are much longer than 120. Padding is used to make the samples of equal length.

\subsection{Experimental Settings}

In the experiments, we performed 5-fold cross-validation in the speaker-independent environment, the proportion of training set, validation set, the test set was set with 8:1:1. The Adam optimizer \cite{kingma2014adam} was used with the initial learning rate of 1e-4 and weight decay set to 1e-8. Moreover, the early stopping was set during training. The batch size was set to 128. We employ unweighted accuracy (UA) and weighted accuracy (WA) as evaluation metrics following the previous studies.  

\subsection{The discussion about FA}
\begin{figure}[htbp]
	\centering
	\includegraphics[width=6cm]{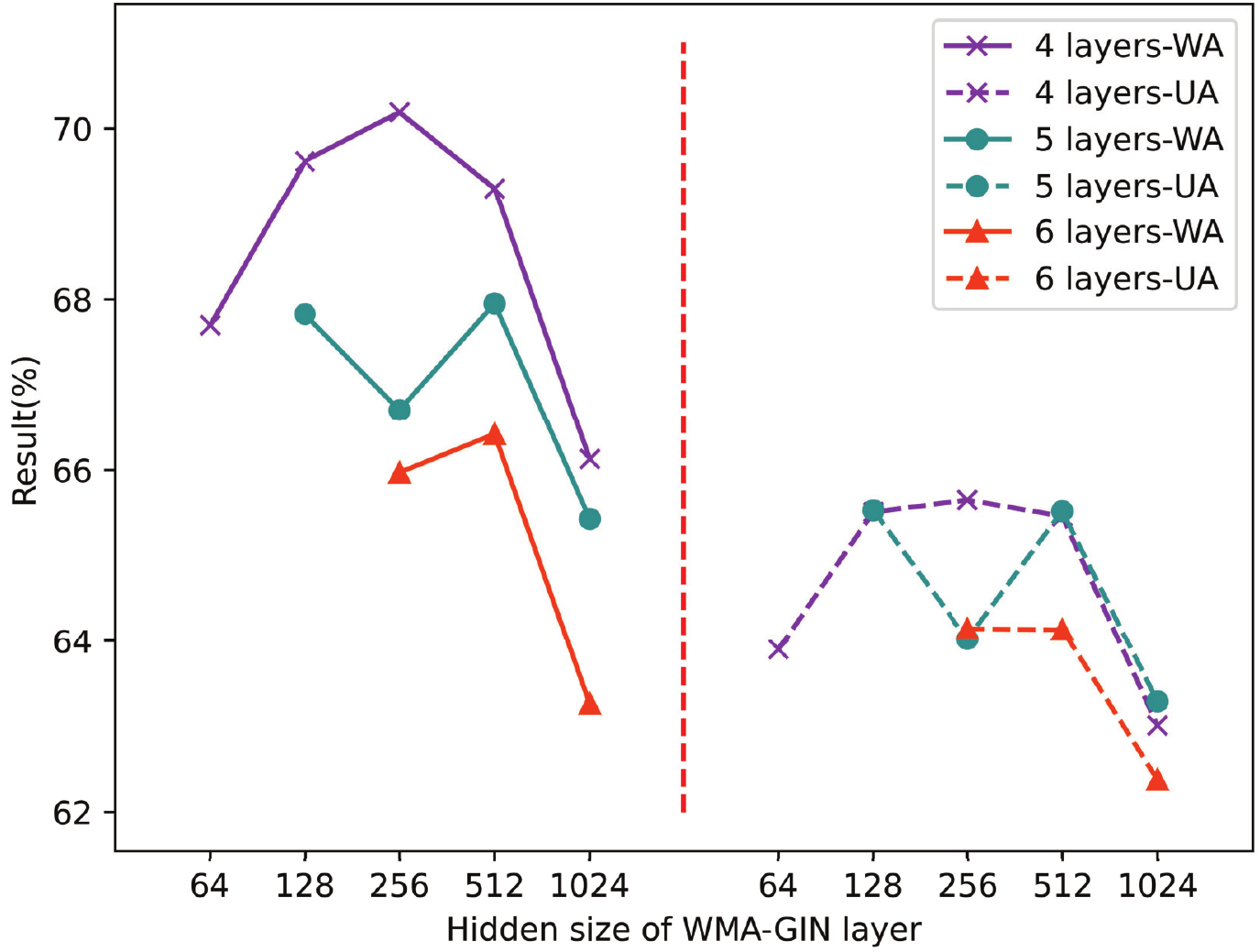}
	\caption{The comparison among the different numbers of WMA-GIN layer along with different hidden sizes}
	\label{hidden size}
\end{figure}

Fig.~\ref{hidden size} shows the UA and WA scores of our proposed method adapting the different numbers of WMA-GIN layer along with different hidden sizes. Alon et.al \cite{alon2020bottleneck} provide the minimal hidden size for different numbers of GIN layers to fit the training data perfectly. Therefore, we set the lower bound of hidden size to 64, 128, and 256 for four, five, and six WMA-GIN layers. The results of WA are all higher than that of UA, and both of them increase first and then decrease with the increase of hidden size. Generally, the more WMA-GIN layer we use, the worse the results are for WA or UA. The model suffers from \(over$-$squashing\) that results in underfitting, which prevents the model from distinguishing between different examples. As a result, we employ the Fully-Adjacent layer in our proposed model once in the following experiments, and adopt the WMA-GIN layers of 4 with the hidden size of 256 according to the result in Fig.\ref{hidden size}.

We further explore which WMA-GIN layer should adopt the FA. Tab.\ref{FA layer} shows three situations: the FA is applied at the last, penultimate, and antepenultimate layer. The model with the FA applied at the antepenultimate layer performs best. Those three situations of applying the FA layers can effectively ease information flow and prevent the \(over$-$squashing\) problem. The model with the FA applied at the antepenultimate layer achieves a better balance between the information interaction with the emotional features extraction. So, in the following experiments, the FA was all applied at the antepenultimate layer in our proposed WMA-GIN model.
\begin{table}[h]
	\caption{The ablation of FA layer}\label{FA layer}
	\centering
	\begin{tabular}{ccc}
		\hline
		FA layer& WA& UA\\
		\hline
		Last FA& 70.67 &65.80\\
		Penultimate layer& 69.68&66.41\\
		Antepenultimate layer& \textbf{72.48}& \textbf{67.72}\\
		\hline
	\end{tabular}
\end{table}

\subsection{The ablation of Weighted Multiple Aggregators}

\begin{table}[h]
	\caption{The ablation of Weighted Multiple Aggregators}\label{Multi-Aggregators}
	\centering
	\begin{tabular}{ccc|c|c}
		\hline
		\multicolumn{3}{c|}{\textbf{Aggregator weight}} & \multicolumn{1}{c|}{\multirow{2}{*}{\textbf{WA(\%)}}} & \multicolumn{1}{c}{\multirow{2}{*}{\textbf{UA(\%)}}}\\ 
		\cline{1-3}
		\multicolumn{1}{c|}{\textbf{softmax}} & \multicolumn{1}{c|}{\textbf{sum}} & \multicolumn{1}{c|}{\textbf{mean}} & \multicolumn{1}{c|}{} & \multicolumn{1}{c}{}\\
		\hline
		\multicolumn{1}{c|}{1/2}     & \multicolumn{1}{c|}{1/4} & 1/4& 68.46& \multicolumn{1}{c}{66.52} \\
		
		\multicolumn{1}{c|}{1/4}     & \multicolumn{1}{c|}{1/2} & 1/4                      & 67.67                                        & \multicolumn{1}{c}{65.31}                   \\
		
		\multicolumn{1}{c|}{1/4}     & \multicolumn{1}{c|}{1/4} & 1/2                      & 67.26                                        & 65.30                                       \\
		
		\multicolumn{1}{c|}{3/5}     & \multicolumn{1}{c|}{1/5} & 1/5                      & 67.88                                        & 65.78                                       \\
		
		\multicolumn{1}{c|}{1/5}     & \multicolumn{1}{c|}{1/5} & 3/5                      & 70.76                                        & 66.24                                       \\
		
		\multicolumn{1}{c|}{1/5}     & \multicolumn{1}{c|}{3/5} & 1/5                      & 69.31                                        & 65.15                                       \\
		
		\multicolumn{1}{c|}{1/3}     & \multicolumn{1}{c|}{1/3} & 1/3                      & \textbf{72.48}                               & \textbf{67.72}                              \\
		
		\multicolumn{1}{c|}{1}       & \multicolumn{1}{c|}{0}   & 0                        & 71.42                                        & 66.59                                       \\
	
		\multicolumn{1}{c|}{0}       & \multicolumn{1}{c|}{1}   & 0                        & 71.26                                        & 66.38                                       \\
		
		\multicolumn{1}{c|}{0}       & \multicolumn{1}{c|}{0}   & 1                        & 69.86                                        & 65.89                                       \\
		\hline
	\end{tabular}
\end{table}

In this section, we explore different weights of three aggregators for aggregating the nodes' information more precisely from different graphs. Furthermore, we also performed experiments with three kinds of single aggregators. As shown in Tab.\ref{Multi-Aggregators}, the model performs the best when three aggregators' weights are equal. That means three aggregators are equally crucial in alleviating information confusion when aggregating node information. As for the single aggregator, the model with the softmax aggregator does the best for the softmax aggregator can allow an asymmetric message passing in the direction of the strongest signal \cite{corso2020principal}. 

\subsection{Comparison with other methods}

As shown in Tab.\ref{graph-based}, the results of GCN \cite{kipf2017semi}, GAT \cite{velivckovic2018graph} are reported by \cite{liu2021graph}, and we reproduced the result of PATCHY-Diff \cite{ying2018hierarchical}. The results of other compared methods are reported in their published papers. Our proposed WMA-GIN outperforms all compared graph-based methods, especially on WA. Compared with LSTM-GIN \cite{liu2021graph} which also adopts GIN structure, WMA-GIN achieves the improvements by 12.1\% and 3.3\% on WA and UA, respectively. Without applying FA, the performance decreases. It proves that the FA layer can effectively alleviate the information compression problem for the SER task. The results in Tab.\ref{graph-based} also indicate that adopting the Multi-stage Loss, assigning different weights to the features at different stages of the network has a positive influence. And the log-Mel spectrogram features are inferior to the LLDs. Furthermore, we also compare six recent advanced methods which do not adopt any GNN structures as shown in Table \ref{non-graph-based}. WMA-GIN utilizing LLDs features outperforms all listed methods on WA.

\begin{table}[h]
	\caption{Comparison with other graph-based methods}\label{graph-based}
	\centering
	\setlength{\tabcolsep}{1.0pt}
	\begin{tabular}{lcccc}
		\hline
		\textbf{Graph-based}& \textbf{Feature sets}&\textbf{Para.}&\textbf{WA(\%)}& \textbf{UA(\%)}\\
		\hline
		GCN \cite{kipf2017semi}&78-LLDs&78K&61.16 &62.21\\
		GAT \cite{velivckovic2018graph}&78-LLDs&-&60.93&62.09\\
		PATCHY-Diff \cite{ying2018hierarchical}&78-LLDs&68K& 63.23& 58.71\\
		GA-GRU \cite{su2020improving}&78-LLDs&- &62.27& 63.80\\
		Amir et.al \cite{shirian2021compact}&35-LLDs&30K& 63.69& 59.87\\
		L-GrIN \cite{shirian2021dynamic}&35-LLDs&92K& 65.50& N/A\\
		LSTM-GIN \cite{liu2021graph}&78-LLDs&0.89M&64.65& 65.53\\
		\hline
		\noindent\textbf{WMA-GIN} & 78-LLDs&0.98M&\textbf{72.48} & \textbf{67.72}\\
		\quad w/o FA & 78-LLDs&-&70.19 & 65.55\\
		\quad w/o Multi-stage Loss &78-LLDs&-&70.25 & 66.47\\
		\noindent\textbf{WMA-GIN}& 128-Log-Mel&0.98M& 65.61&60.26\\
		\hline
	\end{tabular}
\end{table}

\begin{table}[h]
	
	\caption{Comparison with other non-graph-based methods}\label{non-graph-based}
	\centering
	\setlength{\tabcolsep}{4.5pt}
	\begin{tabular}{lcc}
		\hline
		\textbf{Non-Graph-based}&\textbf{WA(\%)}& \textbf{UA(\%)}\\
		\hline
		SegCNN \cite{mao2019deep}&64.53 &62.34\\
		LSTM + CTC \cite{zhao2019attention}& 69.00&67.00\\
		Fine-tune Context Network \cite{xia2021temporal}&66.90& 65.40\\
		Zhong et.al \cite{zhong2020lightweight}&70.39& 71.72\\
		HNSD \cite{cao2021hierarchical}&70.50& 72.50\\
		DAAE+CNN+Attention \cite{gao2021domain} &70.07& 70.67\\
		\hline
		\textbf{WMA-GIN}& \textbf{72.48}& 67.72\\
		\hline
	\end{tabular}
\end{table}

\section{Conclusions}

In this paper, we propose a network based on stacked Weighted Multiple Aggregators Graph Isomorphism Network (WMA-GIN) for SER. The experimental results demonstrate that the WMA  can effectively improve the GIN structure performance. Moreover, the Full-Adjacent (FA) layer is proved to help alleviate \(over$-$squashing\) problem in the SER task. Finally, with the assistance of multi-phase attention (MPA) and multi-stage loss training strategy, WMA-GIN surpasses other graph-based methods and achieves comparable performance to some advanced non-graph-based methods. In the future, we will focus on exploring different graph construction for modeling longer speech utterances in a single graph.
\section{Acknowledgements}

This work is supported by National Natural Science Foundation of China (NSFC) (U1903213), Tianshan Innovation Team Plan Project of Xinjiang (202101642)

\bibliographystyle{IEEEtran}

\bibliography{mybib}


\end{document}